# Non-chiral 1$T$-TiSe$_2$ creates circular dichroism in resonant X-ray diffraction via multipole scattering interference

Hiroki Ueda[1,2,*], Yves Joly[3], and Urs Staub[1,*]

[1]*Center for Photon Science, Forschungsstrasse 111, Paul Scherrer Institute, 5232 Villigen-PSI, Switzerland.*

[2]*Institute of Ion Beam Physics and Materials Research, Helmholtz-Zentrum Dresden-Rossendorf, Dresden 01328, Germany.*

[3]*Université Grenoble Alpes, CNRS, Institut Néel, F-38042 Grenoble, France.*

\*Correspondence authors: hiroki.ueda@psi.ch and urs.staub@psi.ch

## Abstract

Resonant X-ray diffraction (RXD) provides a unique capability to investigate electronically ordered states in matter. Importantly, circular dichroism in RXD can determine the absolute chirality formed by anisotropic multipole arrangement. Here, we demonstrate that the scattering interference between distinct atomic electric multipoles can create circular contrast in resonant RXD even in a non-chiral crystal. These general considerations are applied to 1$T$-TiSe$_2$, where recently reported circular dichroism in RXD was used to claim a chiral charge density wave order. We show that the experimental observations are well reproduced by our *ab initio* calculations based on the centrosymmetric crystal structure. Our results not only illuminate a way for a better understanding of possible chiral electronic order but also propose a general approach to extract buried high-order atomic multipoles via circular contrast in RXD. Besides, we discuss the difference in the circular dichroism of RXD originated by crystal chirality in chiral crystals and the interference effect in non-chiral crystals.

# Introduction

Circularly polarized incident X-ray beams for resonant X-ray diffraction (RXD) provide a unique capability to probe chiral order in matters. A phase shift in scatterings between a co-polarization channel (σ–σ' and π–π') and a cross-polarization channel (σ–π' and π–σ') results in interference of the channels, giving rise to circular contrast, which reverses sign between enantiomers with opposite chirality [1, 2]. Here, σ (π) polarization is defined as being perpendicular (parallel) to the scattering plane. The cross-polarization channels are explicitly zero for non-resonant or resonant charge scattering caused by electric monopole moments (charge), as their atomic scattering factors are isotropic, namely scalers. To obtain circular contrast, one needs tensorial crystal scattering factors that rotate the polarization. This is possible for high-order multipoles resonantly involved in the X-ray scattering process. Within electric dipole-electric dipole (E1E1) transitions, such high-order multipoles can be magnetic dipole and electric quadrupole moments. Namely, chiral order detection by RXD is based on the sensitivity to anisotropic multipoles that align in a chiral manner. This principle also applies to resonant inelastic X-ray scattering, which recently probed chiral phonons via electric quadrupole scatterings [3].

Very recently, Xiao *et al.* reported the emergence of circular contrast on the (0.5 0 2.5) superlattice reflection in RXD intensity from a 1$T$-TiSe$_2$ crystal in the charge density wave (CDW) phase at the Ti $K$ edge. They interpreted the origin of the circular contrast based on a chiral CDW order [4]. Possible chiral CDW order in 1$T$-TiSe$_2$ has extensively been debated for the last fifteen years [5-13], but no consent has been established in the community to date. Thus, the observation of the circular contrast in CDW-origin RXD is stimulating and is potentially a convincing argument for chiral CDW order to settle the debate, considering the powerful capability of RXD with circular polarization as a probe of chiral order.

However, the immediate link between the circular contrast in RXD and a chiral order is unfortunately not that straightforward. Other channels potentially create circular contrast: (1) the birefringence effect [14] and (2) the interference between independent scattering processes [15-18]. These possibilities need to be excluded for the use of circular contrast in support of emergent chirality in 1$T$-TiSe$_2$. Here, we consider the second origin to be relevant to this specific case. Normally, RXD measures the square of an order parameter and, thus, is insensitive to the sign of an order parameter or the phase of a structure factor, e.g., an antiferromagnetic order parameter. However, if more than two scattering processes contribute to RXD, interference between two scattered waves from distinct scattering objects, e.g., electric monopole and magnetic dipole, allows us to extract the sign of an order parameter, which can appear in circular contrast.

Based on parameter-free *ab initio* calculations and symmetry analysis, we show that the circular contrast on the (0.5 0 2.5) reflection in RXD intensity from 1*T*-TiSe$_2$ [4] can be nicely reproduced in the absence of a chiral CDW order, i.e., using the reported non-chiral crystal structure with the centrosymmetric space group *P*–3*c*1 [20]. Therefore, the observation of the circular contrast as such does not support the chiral CDW order. Our results indicate that the circular contrast originates from the interference between different types of atomic electric multipole moments of one of the Ti sites within the electric dipole-electric quadrupole (E1E2) transitions, which substantially contributes to the reflection at the pre-edge of the Ti *K* edge because of the 3*d*-4*p* hybridization on the local space-inversion symmetry breaking on the Ti site.

## Results and discussion

Here, we use the FDMNES (Finite Difference Method Near Edge Structure) code [19], which simulates resonant X-ray spectroscopies, including polarization dependence, e.g., X-ray absorption spectrum (XAS) and RXD. This code is based on the density functional theory methods [19] and is suitable for calculations at the *K* edge, where the core-hole state is far deep in energy and interacts less with unoccupied outer shells. Especially, its capability to simulate site- and transition-specific contributions, i.e., different Ti sites (Ti1 and Ti2 in 1*T*-TiSe$_2$) and E1E1, E1E2, and E2E2 transitions, allows us to identify what contributes to a specific feature such as circular dichroism in RXD.

The (0.5 0 2.5) reflection in RXD appears around the pre-edge of the Ti *K* edge, as shown in Fig. 1(a) {see Fig. 1(d) in Ref. [4] for experimental data}. Here, the main edge is ~4975.25 eV [see Fig. 1(c) for calculated XAS]. A two-dimensional map of the calculated circular contrast ($I_L - I_R$) for incident X-ray energy and azimuthal angle ($\psi$), shown in Fig. 1(b), exhibits clear dichroic signals with a substantial $\psi$ dependence. Here, the origin of $\psi$ is defined when the *a*\* axis is in the scattering plane, as in Ref. [4]. The important point here is that the calculation is based on the *non-chiral* crystal structure and, despite its centrosymmetric symmetry, exhibits the emergence of circular contrast. Energy scans of the RXD intensity at fixed $\psi$ (330°) indicated by a vertical dashed line in Fig. 1(b) [see Figs. 1(d)] show significant circular contrast at the pre-edge, which is in good agreement with the experimental observation.

Detailed analysis revealed that the circular contrast originates from the Ti2 site on the Wyckoff position 6$f$ with the site symmetry .2. mainly through mixed E1E2 transitions [see Fig. 1(d)]. An E2 event between 1$s$ and 3$d$ states intrinsically occurs at the pre-edge because the different screening of the core-hole potential for 3$d$ and 4$p$ states defines their relative energy levels from the 1$s$ state. The local space-inversion symmetry breaking on the Ti2 site allows orbital mixing of 3$d$ and 4$p$ bands. Correspondingly, an E1E2 event between 1$s$ and hybridized 4$p$ states becomes prominent, as commonly observed in XAS at the $K$ edge of 3$d$ transition metals without local space-inversion symmetry [21]. This contribution strongly enhances the pre-edge feature, which is usually very weak for pure E2E2 transitions. E1E2 transitions are sensitive to electromagnetic multipoles breaking space-inversion symmetry with the rank between 1 and 3, i.e., an electric dipole (rank 1), polar toroidal dipole (rank 1), axial toroidal quadrupole (rank 2), magnetic quadrupole (rank 2), electric octupole (rank 3), and polar toroidal octupole (rank 3) [22]. In a non-magnetic material like 1$T$-TiSe$_2$, the magnetic multipoles are explicitly zero and only electric multipoles contribute to RXD.

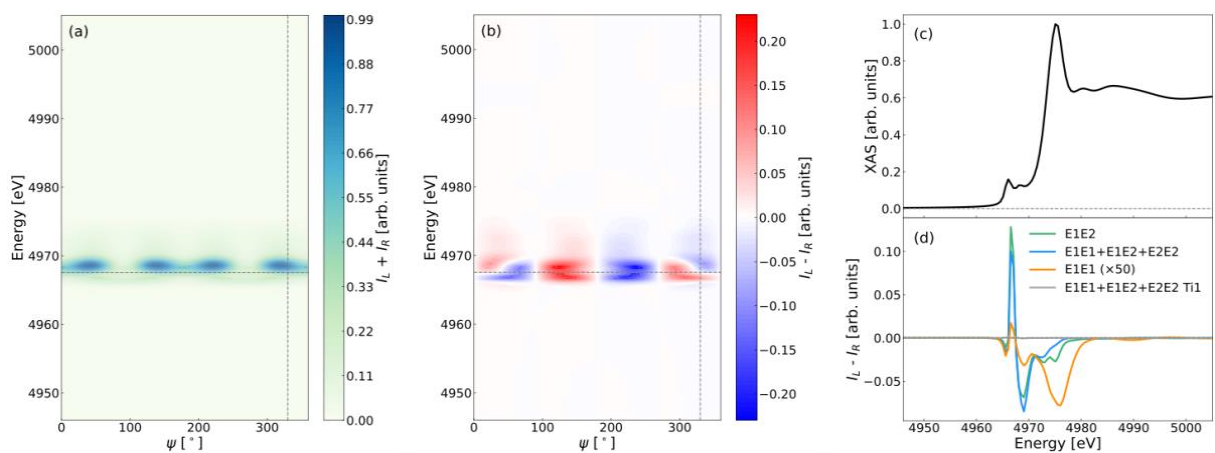

Fig. 1 | (a) Maps of the (0.5 0 2.5) RXD intensity (sum of two circular polarization) and (b) its circular contrast around the Ti $K$ edge. The horizontal and vertical dashed lines indicate the energy (4967.55 eV) and $\psi$ (330°) employed for the $\psi$ dependence (shown in Fig. 3) and energy dependence, respectively. (c) XAS and (d) the energy dependence of the circular contrast in the (0.5 0 2.5) RXD intensity calculated for several conditions, e.g., extracting a specific site [Ti1 site on the Wyckoff position 2$a$ with the site symmetry 32. (gray)] or transition type [E1E1 (orange, multiplied by 50), E1E2 (green), and the sum of E1E1, E1E2, and E2E2 (blue)] contribution. These results include the self-absorption correction (see Fig. 3 for the effect).

Let us formulate the circular contrast signal. A polarization-dependent structure factor $A_{\varepsilon,\varepsilon'}$, where $\varepsilon$ and $\varepsilon'$ represent the polarization state ($\sigma$, $\pi$, R, and L) of the incident and outgoing X-ray beams, respectively, is described as

$$A_{\varepsilon,\varepsilon'} = \sum_{t,l,m} P_{t,l,m}^{\varepsilon,\varepsilon'} X_{t,l,m}(\omega). \qquad (1)$$

Here, $t$ stands for a transition type, e.g., E1E1, E1E2, and E2E2, $l$ and $m$ specify a *crystal multipole*, where $l$ denotes the rank and $-l \leq m \leq l$, and $P_{t,l,m}^{\varepsilon,\varepsilon'}$ is a polarization tensor. The tensorial structure tensor $X_{t,l,m}$ is calculated by summing energy-dependent symmetry-adopted atomic tensors $F_{t,l,m}(\omega)$, i.e.,

$$F_{t,l,m}(\omega) \cong m\omega^2 \int \frac{T_{t,l,m}}{\hbar\omega - (\varepsilon - E_{1s}) + i\frac{\Gamma}{2}} d\varepsilon, \text{ and} \qquad (2)$$

$$X_{t,l,m}(\omega) = \sum_S S[e^{i\mathbf{Q}\cdot\mathbf{r}_1} F_{t,l,m}(\omega)]. \qquad (3)$$

Here, $S$ denotes a symmetry operation connecting equivalent atoms in a unit cell and $T_{t,l,m}$ is an energy-independent atomic tensor. For the Ti2 site, the symmetry operations are $E$ (atom 1), $E \otimes C_{2x}$ (atom 2), $E \otimes C_{3z}$ (atom 3), $I$ (atom 4), $I \otimes C_{2x}$ (atom 5), and $I \otimes C_{3z}$ (atom 6), where $x$ is along the local two-fold axis in the basal plane and $z$ is along [001]. Writing the phase factor $e^{i\mathbf{Q}\cdot\mathbf{r}_1}$ as $a$, where $\mathbf{r}_1$ is the atomic positional vector of atom 1 and $\mathbf{Q}$ is the scattering vector [= (0.5, 0, 2.5) in our case], the structure factor $X_{t,l,m}(\omega)$ is obtained as

$$X_{t,l,m}(\omega) = [(aE + a^*I) - i(E - I)C_{2x} - (a^*E + aI)C_{3z}]F_{t,l,m}(\omega). \qquad (4)$$

The energy-independent part of $X_{t,l,m}(\omega)$ is pure imaginary, and the *energy-dependent* part makes $X_{t,l,m}(\omega)$ complex. This is in contrast to a screw-axis forbidden reflection from a chiral crystal, which gives rise to circular contrast defining the absolute chirality of a crystal structure [2], as the *energy-independent* part of $X_{t,l,m}(\omega)$ can be complex in the chiral crystal case [23, 24]. The circular contrast is given by

$$I_R - I_L = |A_{R,\sigma'}|^2 + |A_{R,\pi'}|^2 - |A_{L,\sigma'}|^2 - |A_{L,\pi'}|^2$$
$$= \left|\sum_{t,l,m} P_{t,l,m}^{R,\sigma'} X_{t,l,m}(\omega)\right|^2 + \left|\sum_{t,l,m} P_{t,l,m}^{R,\pi'} X_{t,l,m}(\omega)\right|^2 - \left|\sum_{t,l,m} P_{t,l,m}^{L,\sigma'} X_{t,l,m}(\omega)\right|^2 -$$
$$\left|\sum_{t,l,m} P_{t,l,m}^{L,\pi'} X_{t,l,m}(\omega)\right|^2. \qquad (5)$$

Since E1E2 transitions dominate the circular contrast, as found in Fig. 1(d), we omit the index $t$ and focus solely on E1E2 transitions. For E1E2 transitions, a polarization tensor holds the relations $P_{l,m}^{R,\sigma'} = P_{l,m}^{L,\sigma'*}$ and $P_{l,m}^{R,\pi'} = P_{l,m}^{L,\pi'*}$. Thus,

$$I_R - I_L = \left|\sum_{l,m} P_{l,m}^{L,\sigma'*} X_{l,m}(\omega)\right|^2 + \left|\sum_{l,m} P_{l,m}^{L,\pi'*} X_{l,m}(\omega)\right|^2 - \left|\sum_{l,m} P_{l,m}^{L,\sigma'} X_{l,m}(\omega)\right|^2$$
$$- \left|\sum_{l,m} P_{l,m}^{L,\pi'} X_{l,m}(\omega)\right|^2$$

$$\begin{aligned}
&= \sum_{l,l',m,m'} \left( P_{l,m}^{L,\sigma'} P_{l',m'}^{L,\sigma'\,*} + P_{l,m}^{L,\pi'} P_{l',m'}^{L,\pi'\,*} - P_{l,m}^{L,\sigma'\,*} P_{l',m'}^{L,\sigma'} - P_{l,m}^{L,\pi'\,*} P_{l',m'}^{L,\pi'} \right) X_{l,m}^*(\omega) X_{l',m'}(\omega) \\
&= 2 \sum_{l \neq l', m \neq m'} \mathrm{Im}\left( P_{l,m}^{L,\sigma'} P_{l',m'}^{L,\sigma'\,*} + P_{l,m}^{L,\pi'} P_{l',m'}^{L,\pi'\,*} \right) \mathrm{Im}\left[ X_{l,m}^*(\omega) X_{l',m'}(\omega) \right] \\
&= 4 \sum_{l<l', m<m'} \mathrm{Im}\left( P_{l,m}^{L,\sigma'} P_{l',m'}^{L,\sigma'\,*} + P_{l,m}^{L,\pi'} P_{l',m'}^{L,\pi'\,*} \right) \mathrm{Im}\left[ X_{l,m}^*(\omega) X_{l',m'}(\omega) \right]. \quad (6)
\end{aligned}$$

Equation (6) is general as long as only a single transition event is relevant and shows that only the interference terms of multipole scatterings ($l \neq l'$ or $m \neq m'$ for *crystal* multipoles) can contribute to the circular contrast regardless of space-inversion symmetry. For a chiral crystal, an *atomic* multipole component $T_{t,l,m}$ can make the energy-independent part of $X_{t,l,m}(\omega)$ complex and create circular contrast in general [23, 24]. In this case, the sign of circular contrast can be independent of photon energy. In contrast, for a centrosymmetric crystal like 1$T$-TiSe$_2$, only the energy-dependent part of $X_{t,l,m}(\omega)$ allows the emergence of circular contrast because none of the energy-independent parts of the $X_{t,l,m}(\omega)$ components are complex. Therefore, circular contrast can appear only through the interference between different *atomic* multipole scatterings from resonators with different energy dependencies. Therefore, their interference is expected to be strongly energy-dependent.

Our calculation demonstrates circular contrast switching at the pre-edge at given $\psi$, as experimentally observed, which supports our interpretation based on the interference effect [see Fig. 1(d)]. The circular contrast switching is in contrast behavior of a chiral crystal [24]. Indeed, while every single multipole contribution term does not create a circular contrast (see the dashed gray line in Fig. 2), it is all through the interference terms, which give rise to finite dichroism in the calculation (see the solid gray curve in Fig. 2). There are seven allowed atomic electric multipoles at the Ti2 site among fifteen atomic electric multipoles that could be detected via E1E2 transitions (i.e., one electric dipole, three axial toroidal quadrupoles, and three electric octupoles). The circular contrast shown by the gray curve in Fig. 2 is observed at photon energies with finite Ti 3$d$ projected density of states, as shown by the blue curve, suggesting the strong involvement of Ti 3$d$ states for the circular contrast.

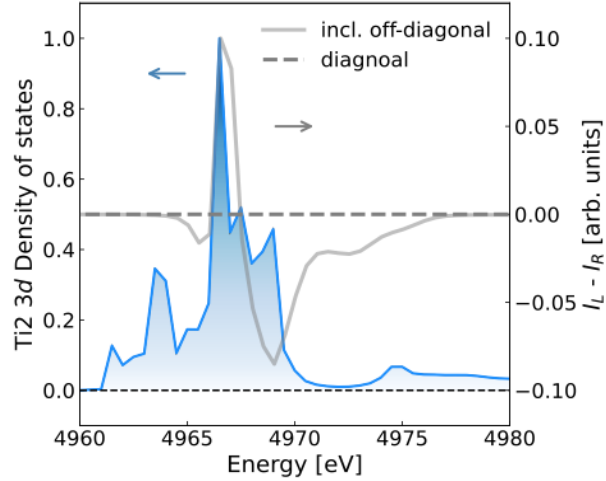

Fig. 2 | Ti 3*d* projected density of states (blue) and the energy dependence of the (0.5 0 2.5) RXD circular contrast with (solid gray curve) or without (dashed gray line) the interference between atomic multipoles.

Finally, we compare our calculations with experimental data shown in Ref. [4]. Figures 3(a) and 3(b) overlay our calculated results at a photon energy of 4967.55 eV onto the experimental data shown in Figs. 2(c) and 2(d) in Ref. [4], i.e., $\psi$ dependence of the (0.5 0 2.5) reflection intensities in RXD with circular polarizations and the normalized circular contrast, respectively. The calculated profiles roughly match the experimental data without any free parameter except for the global amplitude. This interpretation is much simpler and has no free parameters compared to the chiral crystal model used in Ref. [4]. Some deviations at large $\psi$ values, where the experimental data points substantially scatter, might be due to multiple scattering contributions in the experiment. Note that a tiny misalignment from a perfect circular polarization state, i.e., elliptical polarization, can also result in deviation from the nominal curve by calculations [25, 26]. Self-absorption slightly changes the amplitude of the circular contrast, especially at photon energies close to the main edge. Figure 3(c) shows the energy dependence of the circular contrast at $\psi = 330°$. Overall features are reproduced, e.g., the sign flip at the pre-edge and a smaller signal closer to the main edge. As stated above, most signals come from the E1E2 transitions of the Ti2 site. The contributions of the E1E1 transitions and the Ti1 site, which are less than 5%, are nevertheless included in the simulation.

Xiao *et al.* interpreted the observation of the circular contrast as a chiral arrangement of ordered Ti 3*d* $t_{2g}$ and $e_g$ based on the dominant pre-edge feature [4]. Ti 3*d* contribution in their interpretation is also consistent with our calculations, as supported by the Ti 3*d* projected density of states shown in Fig. 2. However, we remark again that our calculations are based on the non-chiral CDW order. Furthermore, our interpretation also circumvents the problem of possible domains, as circular dichroism from opposite chiral domains would cancel. CDW domains are expected to be small compared to the volume probed by the X-ray beams, according to observations by scanning tunneling microscopy experiments on $1T$-TiSe$_2$ [5], as consistent with typical correlation lengths in other CDW systems [27-29]. Thus, it is unlikely that there are coherent chiral domains as large as the X-ray beam used in the experiment [4] (150×70 μm$^2$). Mapping a thin flake sample with RXD using tightly focused circularly polarized X-rays [30] or reconstructing a domain pattern from RXD images with circularly polarized coherent X-rays [31, 32] might be a route to disentangle possible contributions from chiral domains to that from the interference scattering.

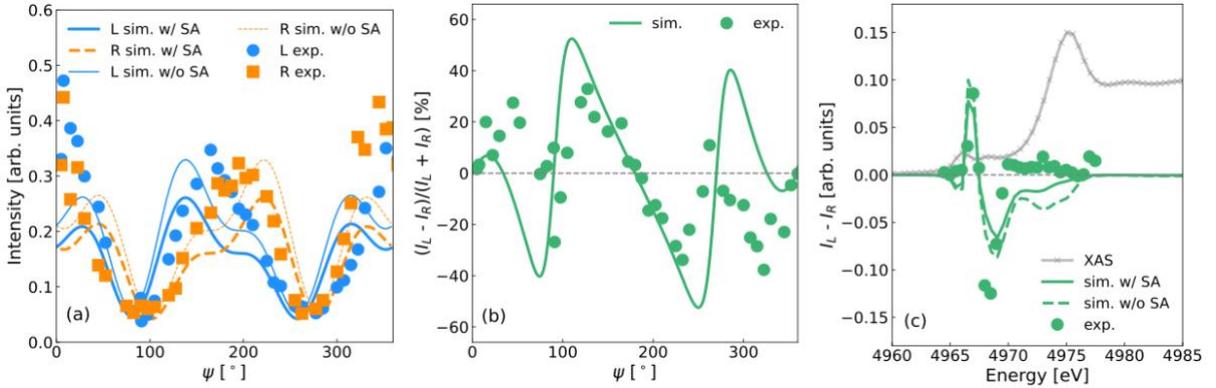

Fig. 3 | Comparison between first-principle calculations (lines) and experimental data (circles, squares). $\psi$ dependence of (a) the (0.5 0 2.5) reflection in RXD with circular polarizations with and without self-absorption (SA) and (b) the normalized circular contrast obtained from (a). (c) The energy dependence of the circular contrast (green) together with XAS shown by the gray curve with × markers [the same data as Fig. 1(c)]. Experimental data are taken from Ref. [4].

## Conclusion

The circular contrast in RXD is a quantity that is important to investigate a chiral order but can be misleading if there are multiple resonators with slightly different energies. The proposed atomic multipole scattering interference in this study can be the origin of or contribute to the circular contrast in RXD in general but is likely more relevant when high-order multipoles contribute to the scattering as they have naturally different resonant energies. This is particularly crucial at the pre-edges of $K$-edge resonances for transition metals because an E1E2 event can be substantial due to orbital mixing between $3d$ and $4p$ via local space-inversion symmetry breaking on the resonant atomic site. Multipole expansion of charge and magnetization densities is a general and robust approach to understanding exotic macroscopic responses in correlated electron systems. Measuring circular contrast in RXD can be a route to extract hidden higher-order multipoles from dominant lower-rank multipole contributions, even without a super-lattice reflection solely formed by higher-order multipole order. Our results do not directly deny the chiral CDW order in 1$T$-TiSe$_2$ but show that a measurement of an inversion of circular dichroism between chiral domains is required to verify a chiral order. Here, we show that the experimentally observed circular contrast in RXD from the CDW phase of 1$T$-TiSe$_2$ at the Ti $K$ edge can be nicely explained by a "simple" parameter-free model based on the first-principle approach in the centrosymmetric symmetry without the need for having a chiral CDW order.

## References

1. C. Sutter, G. Grübel, C. Vettier, F. de Bergevin, A. Stunault, D. Gibbs, and C. Giles, Helicity of magnetic domains in holmium studied with circularly polarized x rays. *Phys. Rev. B* **55**, 954-959 (1997). DOI: https://doi.org/10.1103/PhysRevB.55.954
2. Y. Tanaka, T. Takeuchi, S. W. Lovesey, K. S. Knight, A. Chainani, Y. Takata, M. Oura, Y. Senba, H. Ohashi, and S. Shin, Right handed or left handed? Forbidden x-ray diffraction reveals chirality. *Phys. Rev. Lett.* **100**, 145502 (2008). DOI: https://doi.org/10.1103/PhysRevLett.100.145502
3. H. Ueda, M. García-Fernández, S. Agrestini, C. P. Romao, J. van den Brink, N. A. Spaldin, K.-J. Zhou, and U. Staub, Chiral phonons in quartz probed by X-rays. *Nature* **618**, 946-950 (2023). DOI: https://doi.org/10.1038/s41586-023-06016-5
4. Q. Xiao, O. Janson, S. Francoual, Q. Qiu, Q. Li, S. Zhang, W. Xie, P. Bereciartua, J. van den Brink, J. van Wezel, and Y. Peng, Observation of circular dichroism induced by electronic chirality. *Phys. Rev. Lett.* **133**, 126402 (2024). DOI: https://doi.org/10.1103/PhysRevLett.133.126402
5. J. Ishioka, Y. H. Liu, K. Shimatake, T. Kurosawa, K. Ichimura, Y. Toda, M. Oda, and S. Tanda, Chiral charge-density waves. *Phys. Rev. Lett.* **105**, 176401 (2010). DOI: https://doi.org/10.1103/PhysRevLett.105.176401
6. J.-P. Castellan, S. Rosenkranz, R. Osborn, Q. Li, K. E. Gray, X. Luo, U. Welp, G. Karapetrov, J. P. C. Ruff, and J. van Wezel, Chiral phase transition in charge ordered 1$T$-TiSe$_2$. *Phys. Rev. Lett.* **110**, 196404 (2013). DOI: https://doi.org/10.1103/PhysRevLett.110.196404

7. B. Hildebrand, T. Jaouen, M.-L. Mottas, G. Monney, C. Barreteau, E. Giannini, D. R. Bowler, and P. Aebi, Local real-space view of the achiral 1*T*-TiSe$_2$ 2×2×2 charge density wave. *Phys. Rev. Lett.* **120**, 136404 (2018). DOI: https://doi.org/10.1103/PhysRevLett.120.136404

8. M.-K. Lin, J. A. Hlevyack, P. Chen, R.-Y. Liu, and T.-C. Chiang, Comment on "Chiral phase transition in charge ordered 1*T*-TiSe$_2$". *Phys. Rev. Lett.* **122**, 229701 (2019). DOI: https://doi.org/10.1103/PhysRevLett.122.229701

9. S.-Y. Xu, Q. Ma, A. Kogar, A. Zong, A. M. Mier Valdivia, T. H. Dinh, S.-M. Huang, B. Singh, C.-H. Hsu, T.-R. Chang, J. P. C. Ruff, K. Watanabe, T. Taniguchi, H. Lin, G. Karapetrov, D. Xiao, P. Jarillo-Herrero, and N. Gedik, Spontaneous gyrotropic electronic order in a transition-metal dichalcogenide. *Nature* **578**, 545-549 (2020). DOI: https://doi.org/10.1038/s41586-020-2011-8

10. H. Ueda, M. Porer, J. R. L. Mardegan, S. Parchenko, N. Gurung, F. Fabrizi, M. Ramakrishnan, L. Boie, M. J. Neugebauer, B. Burganov, M. Burian, S. L. Johnson, K. Rossnagel, and U. Staub, Correlation between electronic and structural orders in 1*T*-TiSe$_2$. *Phys. Rev. Res.* **3**, L022003 (2021). DOI: https://doi.org/10.1103/PhysRevResearch.3.L022003

11. R. Zhang, W. Ruan, J. Yu, L. Gao, H. Berger, L. Forró, K. Watanabe, T. Taniguchi, A. Ranjbar, R. V. Belosludov, T. D. Kühne, M. S. Bahramy, and X. Xi, Second-harmonic generation in atomically thin 1*T*-TiSe$_2$ and its possible origin from charge density wave transitions. *Phys. Rev. B* **105**, 085409 (2022). DOI: https://doi.org/10.1103/PhysRevB.105.085409

12. K. Kim, H.-W. J. Kim, S. Ha, H. Kim, J.-K. Kim, J. Kim, J. Kwon, J. Seol, S. Jung, C. Kim, D. Ishikawa, T. Manjo, H. Fukui, A. Q. R. Baron, A. Alatas, A. Said, M. Merz, T. L. Tacon, J. M. Bok, K.-S. Kim, and B. J. Kim, Origin of the chiral charge density wave in transition-metal dichalcogenide. *Nat. Phys.* published only online. (2024). DOI: https://www.nature.com/articles/s41567-024-02668-w

13. H. Kim, K.-H. Jin, and H. W. Yeom, Electronically seamless domain wall of chiral charge density wave in 1*T*-TiSe$_2$. *Nano Lett.* Published only online. DOI: https://doi.org/10.1021/acs.nanolett.4c03970

14. Y. Joly, S. P. Collins, S. Grenier, H. C. N. Tolentino, and M. De Santis, Birefringence and polarization rotation in resonant x-ray diffraction. *Phys. Rev. B* **86**, 220101(R) (2012). DOI: https://doi.org/10.1103/PhysRevB.86.220101

15. S. Tardif, S. Takeshita, H. Ohsumi, J. Yamaura, D. Okuyama, Z. Hiroi, M. Takata, and T. Arima, All-in-all-out magnetic domains: x-ray diffraction imaging and magnetic field control. *Phys. Rev. Lett.* **114**, 147205 (2015). DOI: https://doi.org/10.1103/PhysRevLett.114.147205

16. H. Ueda, Y. Tanaka, Y. Wakabayashi, and T. Kimura, Observation of collinear antiferromagnetic domains making use of the circular dichroic charge-magnetic interference effect of resonant x-ray diffraction. *Phys. Rev. B* **98**, 134415 (2018). DOI: https://doi.org/10.1103/PhysRevB.98.134415

17. R. Misawa, H. Ueda, K. Kimura, Y. Tanaka, and T. Kimura, Chirality and magnetic quadrupole order in $Pb(TiO)Cu_4(PO_4)_4$ probed by interference scattering in resonant x-ray diffraction. *Phys. Rev. B* **103**, 174409 (2021). DOI: https://doi.org/10.1103/PhysRevB.103.174409

18. R. Misawa, K. Arakawa, T. Yoshioka, H. Ueda, F. Iga, K. Tamasaku, Y. Tanaka, and T. Kimura, Resonant x-ray diffraction study using circularly polarized x rays on antiferromagnetic $TbB_4$. *Phys. Rev. B* **108**, 134433 (2023). DOI: https://doi.org/10.1103/PhysRevB.108.134433

19. O. Bunau and Y. Joly, Self-consistent aspects of x-ray absorption calculations. *J. Phys.: Condens. Matter* **21**, 345501 (2009). DOI: 10.1088/0953-8984/21/34/345501

20. S. Kitou, A. Nakano, S. Kobayashi, K. Sugawara, N. Katayama, N. Maejima, A. Machida, T. Watanuki, K. Ichimura, S. Tanda, T. Nakamura, and H. Sawa, Effect of Cu intercalation and pressure on excitonic interaction in $1T$-$TiSe_2$. *Phys. Rev. B* **99**, 104109 (2019). DOI: https://doi.org/10.1103/PhysRevB.99.104109

21. D. Cabaret, A. Bordage, A. Juhin, M. Arfaoui, and E. Gaudry, First-principles calculations of X-ray absorption spectra at the $K$-edge of $3d$ transition metals: an electronic structure analysis of the pre-edge. *Phys. Chem. Chem. Phys.* **12**, 5619-5633 (2010). DOI: https://doi.org/10.1039/B926499J

22. S. Di Matteo, Resonant x-ray diffraction: multipole interpretation. *J. Phys. D: Appl. Phys.* **45**, 163001 (2012). DOI: 10.1088/0022-3727/45/16/163001

23. Y. Tanaka, T. Kojima, Y. Takata, A. Chainani, S. W. Lovesey, K. S. Knight, T. Takeuchi, M. Oura, Y. Senba, H. Ohashi, and S. Shin, Determination of structural chirality of berlinite and quartz using resonant x-ray diffraction with circularly polarized x-rays. *Phys. Rev. B* **81**, 144104 (2010). DOI: https://doi.org/10.1103/PhysRevB.81.144104

24. H. Ueda, E. Skoropata, M. Burian, V. Ukleev, G. S. Perren, L. Leroy, J. Zaccaro, and U. Staub, Conical spin order with chiral quadrupole helix in CsCuCl$_3$. *Phys. Rev. B* **105**, 144408 (2022). DOI: https://doi.org/10.1103/PhysRevB.105.144408

25. J. Igarashi, and M. Takahashi, Resonant x-ray scattering from chiral materials: α-quartz and α-berlinite. *Phys. Rev. B* **86**, 104116 (2012). DOI: https://doi.org/10.1103/PhysRevB.86.104116

26. Y. Joly, Y. Tanaka, D. Cabaret, and S. P. Collins, Chirality, birefringence, and polarization effects in *α*-quartz studied by resonant elastic x-ray scattering. *Phys. Rev. B* **89**, 224108 (2014). DOI: https://doi.org/10.1103/PhysRevB.89.224108

27. J.-D. Su, A. R. Sandy, J. Mohanty, O. G. Shpyrko, and M. Sutton, Collective pinning dynamics of charge-density waves in 1*T*-TaS$_2$. *Phys. Rev. B* **86**, 205105 (2012). DOI: https://doi.org/10.1103/PhysRevB.86.205105

28. X. M. Chen, C. Mazzoli, Y. Cao, V. Thampy, A. M. Barbour, W. Hu, M. Lu, T. A. Assefa, H. Miao, G. Fabbris, G. D. Gu, J. M. Tranquada, M. P. M. Dean, S. B. Wilkins, and I. K. Robinson, Charge density wave memory in a cuprate superconductor. *Nat. Commun.* **10**, 1435 (2019). https://doi.org/10.1038/s41467-019-09433-1

29. Z. Shi, S. J. Kuhn, F. Flicker, T. Helm, J. Lee, W. Steinhardt, S. Dissanayake, D. Graf, J. Ruff, G. Fabbris, D. Haskel, and S. Haravifard, Incommensurate two-dimensional checkerboard charge density wave in the low-dimensional superconductor Ta$_4$Pd$_3$Te$_{16}$. *Phys. Rev. Res.* **2**, 042042(R) (2020).

30. T. Usui, Y. Tanaka, H. Nakajima, M. Taguchi, A. Chainani, M. Oura, S. Shin, N. Katayama, H. Sawa, Y. Wakabayashi, and T. Kimura, Observation of quadrupole helix chirality and its domain structure in DyFe$_3$(BO$_3$)$_4$. *Nat. Mater.* **13**, 611-618 (2014). DOI: https://doi.org/10.1038/nmat3942

31. A. Tripathi, J. Mohanty, S. H. Dietze, and I. McNulty, Dichroic coherent diffractive imaging. *Proc. Natl. Acad. Sci.* **108**, 13393-13398 (2011). DOI: https://doi.org/10.1073/pnas.1104304108

32. M. G. Kim, H. Miao, B. Gao, S.-W. Cheong, C. Mazzoli, A. Barbour, W. Hu, S. B. Wilkins, I. K. Robinson, M. P. M. Dean, and V. Kiryukhin, Imaging antiferromagnetic antiphase domain boundaries using magnetic Bragg diffraction phase contrast. *Nat. Commun.* **9**, 5013 (2018). DOI: https://doi.org/10.1038/s41467-018-07350-3

33. H. Ueda, Y. Joly, and U. Staub, PSI Data Repository https://doi.org/10.16907/d77e95b6-144f-4a61-9259-3b04888c718b

## Competing interests

The authors declare no competing interests.

## Data availability

Calculated data are accessible from the PSI Public Data Repository [33].